# The mutual rectification effect of electric currents induced by electromagnetic waves in materials with non-additive energy spectrum:
# the quantum kinetic equation approach.

*V. I. Konchenkov, S. V. Kryuchkov, D. V. Zav'yalov*

Volgograd State Pedagogical University

400131, Russia, Volgograd, Lenina av., 27

sed@fizmat.vspu.ru

The effect of direct current component appearance in material with non-additive energy spectrum in the case when two electromagnetic waves with mutually transverse planes of polarization are incident normally on the surface of the sample is studied on the base of analysis of quantum kinetic equation. The cases of electron scattering on optical and acoustic phonons are considered. The first non-vanishing approximation by electromagnetic waves field strengths is obtained. It is shown that in concerned situation the decisive condition of direct current appearance is sufficient inelasticity of scattering.

Charge carriers energy spectrum non-additivity is the cause of interdependence between motions along mutually perpendicular axes. This interdependence may lead to symmetry breaking of material transport properties. Among the other effects which can be found in materials because of their charge carriers spectrum non-additivity, there are the effects of alternating currents rectification. The appearance of continuous current in direction along which there are no constant electric fields only by interference between fields of electromagnetic waves incident to the surface of the sample is one of examples of the same effects. These effects can be used to experimental studying of different properties of materials as well as some properties of radiation such as presence of second harmonic of radiation, phase shift and other.

Progress of nanotechnology in recent years made possible to create new materials with unusual properties. One of such materials is graphene which first was created in 2004 [1]. Graphene is two-dimensional carbon allotrope. Now there are several modifications of graphene: exfoliation graphene on substrate of $SiO_2$ [1, 2], epitaxial graphene on $SiC$ wafer [2, 3], bilayer graphene [2, 4, 5, 6]. This new material has a lot of remarkable mechanical and electronic properties and so more scientists' attention concentrates on studying them [2]. Many researchers suppose that graphene can become a new base of electronics. There are experimental models of field effect transistors on the base of graphene already [7, 8]. Energy spectrum of graphene is non-additive and non-quadratic [9], so the theoretical results of studying the influence of charge carriers spectrum non-additivity on kinetic properties of materials can be experimentally proved in graphene.

As stated above the non-additivity of energy spectrum can lead to occurrence so named effects of currents rectification. In [10] the effect of appearance of continuous current in direction perpendicular to drawing constant electric field was studied theoretically in semi-classical situation in a case when elliptically polarized electromagnetic wave is incident normally to the surface of graphene on a substrate of $SiO_2$. Energy spectrum of this graphene modification has peculiarity in $|\mathbf{p}|=0$ so it is impossible to solve correctly the problem even in a limit of small field strengths. In [10] distribution function was obtained from Boltzmann equation with the collision term in approximation of constant relaxation time. This approach does not give any information about relative contribution of different scattering mechanism. So in [11] studying of the problem similar to the problem in [10] was attempted with using semi-classical Monte Carlo simulation. In this paper graphene on $SiC$ substrate considered because an energy spectrum of this modification of graphene has a gap with halfwidth $\Delta \approx 0.26 eV$ so it is possible to use one-band approximation. This spectrum also has not any peculiarities at small momentum values therefore analytical and numerical investigations are simplified. In [11] it is shown that direct current component appears only if electron scattering in material is inelastic. In the case of graphene scattering by optical phonons is sufficiently inelastic whereas scattering by acoustic phonons can be considered as elastic [12].

In [13] direct current is generated in single layer graphene by circularly polarized terahertz laser radiation at normal as well as at oblique incidence and changes its sign upon reversing the radiation helicity. In the case of wave oblique incidence the appearance of direct current is caused by the circular photon drag effect in the interior of graphene sheet. But in the case of normal incidence the effect is due to the influence of the sample edges which reduce the equivalence of directions in the samle and result in an asymmetric scattering of carriers driven by the radiation field. This effect has the second order by strengths of fields of the wave while the effect considered in [10, 11] has the third order.

Paper [14] is devoted to studying the mutual rectification effect of alternating currents induced by two electromagnetic waves with perpendicular each other planes of polarization and different frequencies which are incident normally to the surface of graphene. It was shown that direct current appears along the direction of electric-field vector of wave which frequency is two times more than the frequency of another wave. In [15] the effect of direct current appearance is studied in the case when two electromagnetic waves with mutually perpendicular planes of polarization are incident normally to the surface of graphene and constant magnetic field directed perpendicularly to the surface of the sample is presented. It was shown that in the case of presence of magnetic field direct current occurs not only along the direction of electric-field vector of wave with a frequency twice as much as the frequency of another wave but along the direction of electric-field vector of the second wave too. The last current component is proportional to the strength of magnetic field. In [14, 15] the approximation of constant relaxation time was used to obtain the distribution function. As stated above, this approximation does not describe the microscopic processes of electron scattering. So it is interesting to study relative contribution to the direct current component by different mechanism of scattering and additional conditions of continuous current appearance.

To take in account different mechanism of relaxation in materials with non-additive spectrum in this paper method of quantum kinetic equation is used. This method in application to problems of semiconductor physics was developed in [16 - 21].

We shall consider two-dimensional sample with model energy spectrum

$$\varepsilon(\mathbf{p}) = A + B\mathbf{p}^2 + C\mathbf{p}^4, \qquad (1)$$

where $\mathbf{p} = (p_x, p_y)$ - is quasi-momentum vector of electron. The non-additivity of this spectrum is provided by presence of term proportional to $\mathbf{p}^4$. Each non-additive spectrum can be presented in form (1) at sufficiently small values of $|\mathbf{p}|$. Specifically, the spectrum of bilayer graphene is well-described [2, 4, 6] by (1) at the temperatures about $50 K$. In this case $A = \Delta$, $B = -2\Delta v_f^2 / t_\perp^2$, $C = v_f^4 / (2t_\perp^2 \Delta)$, where $t_\perp \approx 0.35 eV$ - is overlap integral between layers of graphene, $v_f \approx 10^8 cm/s$ - is Fermi velocity, $\Delta$ - is halfwidth of bandgap. To use this approximated spectrum for graphene one must satisfy the condition $\Delta \ll t_\perp$.

We shall study the case when two electromagnetic waves with mutually perpendicular planes of polarization are incident normally on the surface of the sample. The geometry of problem is shown on fig. 1. The electric field strength of waves is

$$\mathbf{E} = (E_{10} \sin(\omega_1 t),\ E_{20} \sin(\omega_2 t + \gamma)). \qquad (2)$$

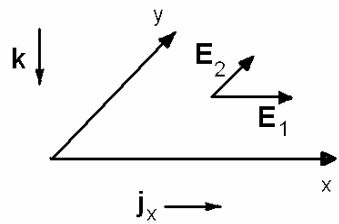

Figure 1. Problem geometry.

Quantum kinetic equation is [16, 21]:

$$\frac{\partial \varphi_\mathbf{p}(t)}{\partial t} = \sum_\mathbf{q} \frac{|C_\mathbf{q}|^2}{\hbar^2} \int_{-\infty}^{t} dt' (Q(\mathbf{p}-\mathbf{q},\mathbf{p}) - Q(\mathbf{p},\mathbf{p}+\mathbf{q})), \qquad (3)$$

$Q(\mathbf{p}-\mathbf{q},\mathbf{p}) =$

$$= \exp\left(-\frac{i}{\hbar}\int_{t'}^{t} dt'' \left(\varepsilon\left(\mathbf{p}-\mathbf{q}-\frac{e}{c}\mathbf{A}(t'')\right)-\varepsilon\left(\mathbf{p}-\frac{e}{c}\mathbf{A}(t'')\right)+\hbar\omega_{\mathbf{q}}\right)\right)\left(\varphi_{\mathbf{p}-\mathbf{q}}(1-\varphi_{\mathbf{p}})(N_{\mathbf{q}}+1)-\varphi_{\mathbf{p}}(1-\varphi_{\mathbf{p}-\mathbf{q}})N_{\mathbf{q}}\right)+$$

$$+ \exp\left(-\frac{i}{\hbar}\int_{t'}^{t} dt'' \left(\varepsilon\left(\mathbf{p}-\mathbf{q}-\frac{e}{c}\mathbf{A}(t'')\right)-\varepsilon\left(\mathbf{p}-\frac{e}{c}\mathbf{A}(t'')\right)-\hbar\omega_{\mathbf{q}}\right)\right)\left(\varphi_{\mathbf{p}-\mathbf{q}}(1-\varphi_{\mathbf{p}})N_{\mathbf{q}}-\varphi_{\mathbf{p}}(1-\varphi_{\mathbf{p}-\mathbf{q}})(N_{\mathbf{q}}+1)\right);$$

$\mathbf{A} = -c\int_{}^{t''}\mathbf{E}d\tau'$ - is a vector potential of electric field of waves, $\mathbf{p}$ - is a quasi-momentum of electron, $\mathbf{q}$ - is a quasi-momentum of phonon, $\omega_{\mathbf{q}}$ - is phonon cyclic frequency, $\varphi(\mathbf{p},t)$ - is an electron distribution function, $N_{\mathbf{q}}$ - is a phonon distribution function, $C_{\mathbf{q}}$ - is a constant of electron-phonon interaction, $c$ - is the speed of light, $\hbar$ - is the Planck constant, $e$ - is absolute value of elementary charge. In deriving this equation it was assumed that electron-phonon interaction is week and phonon gas is equilibrium so $N_{\mathbf{q}} = \frac{1}{\exp(\hbar\omega_{\mathbf{q}}/kT)-1}$. In this paper we shall consider nondegenerate electron gas. Let us consider two cases of electron scattering: electron scattering by nonpolar acoustic phonons and by optical phonons.

In case of scattering by nonpolar acoustic phonons [19, 20, 22] $\hbar\omega_{\mathbf{q}} \ll kT$, so $N_{\mathbf{q}} = \frac{1}{\exp(\hbar\omega_{\mathbf{q}}/kT)-1} \approx \frac{kT}{\hbar\omega_{\mathbf{q}}}$, where $k$ - is a Boltzmann constant, $T$ - is temperature,

$$Q_{ac}(\mathbf{p}-\mathbf{q},\mathbf{p}) = \frac{2kT}{\hbar\omega_{\mathbf{q}}} \exp\left(-\frac{i}{\hbar}\int_{t'}^{t} dt'' \left(\varepsilon\left(\mathbf{p}-\mathbf{q}-\frac{e}{c}\mathbf{A}(t'')\right)-\varepsilon\left(\mathbf{p}-\frac{e}{c}\mathbf{A}(t'')\right)\right)\right)(\varphi_{\mathbf{p}-\mathbf{q}}-\varphi_{\mathbf{p}}), \quad (4)$$

$\left|C_{\mathbf{q}}^{ac}\right|^2 = \frac{D_a^2 q\hbar}{2\rho s L^2}$ [19, 22], where $D_a$ - is a constant of deformation potential, $s$ - is a speed of sound, $L$ - is a linear dimension of a basic area of crystal, $\rho$ - is a surface density of crystal.

In case of scattering by optical phonons $\hbar\omega_{\mathbf{q}} \gg kT$ so $N_{\mathbf{q}} \ll 1$

$$Q_{opt}(\mathbf{p}-\mathbf{q},\mathbf{p}) = \exp\left(-\frac{i}{\hbar}\int_{t'}^{t} dt'' \left(\varepsilon\left(\mathbf{p}-\mathbf{q}-\frac{e}{c}\mathbf{A}(t'')\right)-\varepsilon\left(\mathbf{p}-\frac{e}{c}\mathbf{A}(t'')\right)\right)+\hbar\omega_{\mathbf{q}}\right)\varphi_{\mathbf{p}-\mathbf{q}} -$$

$$- \exp\left(-\frac{i}{\hbar}\int_{t'}^{t} dt'' \left(\varepsilon\left(\mathbf{p}-\mathbf{q}-\frac{e}{c}\mathbf{A}(t'')\right)-\varepsilon\left(\mathbf{p}-\frac{e}{c}\mathbf{A}(t'')\right)\right)-\hbar\omega_{\mathbf{q}}\right)\varphi_{\mathbf{p}}, \quad (5)$$

$\left|C_{\mathbf{q}}^{opt}\right|^2 = \dfrac{\hbar D_o^2}{2\rho\omega_0 L^2}$ [19, 22], where $D_o$ - is the constant of electron interaction with optical phonons, $\omega_o$ - is phonon cyclic frequency (we assume optical phonons was dispersionless).

Let us consider next expression separately

$$I(\mathbf{p}-\mathbf{q},\mathbf{p}) = \dfrac{1}{\hbar}\int_{t'}^{t} dt'' \left(\varepsilon\left(\mathbf{p}-\mathbf{q}-\dfrac{e}{c}\mathbf{A}(t'')\right) - \varepsilon\left(\mathbf{p}-\dfrac{e}{c}\mathbf{A}(t'')\right)\right). \tag{6}$$

We introduce new variables: $\mathbf{p} \to P\mathbf{p}$, $\mathbf{q} \to P\mathbf{q}$, $t \to t/\omega_1$, where $P$ - is some constant with momentum dimension, and enter a new table of symbols: $b = \omega_2/\omega_1$, $\omega_\mathbf{q}' = \omega_\mathbf{q}/\omega_1$, $F_1 = eE_{10}/(P\omega_1)$, $F_2 = eE_{20}/(P\omega_2)$, $B_1 = BP^2/(\hbar\omega_1)$, $C_1 = CP^2/B$. We obtain

$$I(\mathbf{p}-\mathbf{q},\mathbf{p}) = \int_{t'}^{t} dt'' (K_0 + K_1 \cos t + K_2 \cos 2t + K_3 \cos 3t + K_4 \cos(bt+\gamma) + K_5 \cos(2bt+2\gamma) +$$

$$+ K_6 \cos(3bt+3\gamma) + K_7 \cos((2b-1)t+2\gamma) + K_8 \cos((2b+1)t+2\gamma) +$$

$$+ K_9 \cos((b-1)t+\gamma) + K_{10} \cos((b+1)t+\gamma) + K_{11} \cos((b-2)t+\gamma) + K_{12} \cos((b+2)t+\gamma)); \tag{7}$$

Quantities $K_i$ are presented in Appendix.

It is clear from expression (7) that it is necessary to consider four different cases of frequency ratio: $b=1$, $b=2$, $b=1/2$ and $b$ is any other value which is not equal to the first three. The direct estimations, performed for all these cases, show that direct current component arises only for $b=1/2$ so all further calculations will give for this case only. Let us do integration in (7) and substitute the result in (4) and (5). Combining (4) with (3) we obtain quantum kinetic equation in the case of electron scattering on acoustic phonons:

$$\dfrac{\partial \varphi_\mathbf{p}^{ac}(t)}{\partial t} =$$

$$= \sum_\mathbf{q} \xi_{ac} \operatorname{Re} \int_{-\infty}^{0} dt' \sum_{n_1'\ldots n_{12}'}\sum_{n_1\ldots n_{12}} [\{S_{nn'}(\mathbf{p}-\mathbf{q},\mathbf{p})\exp(i(\alpha_0(\mathbf{p}-\mathbf{q},\mathbf{p})+\alpha_1)t' + i\alpha_2 t + i\alpha_3 \gamma)(\varphi_{\mathbf{p}-\mathbf{q}}^{ac} - \varphi_\mathbf{p}^{ac})\} -$$

$$- \{\mathbf{p} \to \mathbf{p}+\mathbf{q}\}], \tag{8}$$

where $\xi_{ac} = \dfrac{2}{(\hbar\omega_1)^2}\dfrac{kT}{\hbar\omega_\mathbf{q}}\left|C_\mathbf{q}^{ac}\right|^2 = \dfrac{1}{(\hbar\omega_1)^2}\dfrac{D_a^2 kT}{\rho s^2 L^2}$.

Similarly, by combining (5) with (3) we receive quantum kinetic equation in the case of scattering on optical phonons:

$$\frac{\partial \varphi_{\mathbf{p}}^{opt}}{\partial t} =$$

$$= \sum_{\mathbf{q}} \xi_{opt} \operatorname{Re} \int_{-\infty}^{0} dt' \sum_{n'...n_{12}'} \sum_{n_1...n_{12}} \left[ \left\{ \left( S_{nn'}(\mathbf{p}-\mathbf{q},\mathbf{p}) \exp\left(i(\alpha_0(\mathbf{p}-\mathbf{q},\mathbf{p})+\alpha_1+\omega_{\mathbf{q}}')t'+i\alpha_2 t+i\alpha_3\gamma\right) \varphi_{\mathbf{p}-\mathbf{q}}^{opt} \right. \right.$$

$$\left. \left. - S_{nn'}(\mathbf{p}-\mathbf{q},\mathbf{p}) \exp\left(i(\alpha_0(\mathbf{p}-\mathbf{q},\mathbf{p})+\alpha_1-\omega_{\mathbf{q}}')t'+i\alpha_2 t+i\alpha_3\gamma\right) \varphi_{\mathbf{p}}^{opt} \right\} - \{\mathbf{p} \to \mathbf{p}+\mathbf{q}\} \right] \quad (9)$$

where $\xi_{opt} = \frac{1}{2(\hbar\omega_1)^2} \frac{\hbar D_o^2}{\rho \omega_0 L^2}$.

Next denotations were introduced:

$$S_{nn'}(\mathbf{p}-\mathbf{q},\mathbf{p}) = J_{n_1'}(K_1) J_{n_2'}\left(\frac{K_2}{2}\right) J_{n_3'}\left(\frac{K_3}{3}\right) J_{n_4'}\left(\frac{K_4}{b}\right) J_{n_5'}(K_5) J_{n_6'}\left(\frac{2K_6}{3}\right) J_{n_8'}\left(\frac{K_8}{2}\right) \cdot$$

$$\cdot J_{n_9'}(-2K_9) J_{n_{10}'}\left(\frac{2K_{10}}{3}\right) J_{n_{11}'}\left(-\frac{2K_{11}}{3}\right) J_{n_{12}'}\left(\frac{2K_{12}}{5}\right) J_{n_1}(K_1) J_{n_2}\left(\frac{K_2}{2}\right) J_{n_3}\left(\frac{K_3}{3}\right) J_{n_4}\left(\frac{K_4}{b}\right) \cdot$$

$$\cdot J_{n_5}(K_5) J_{n_6}\left(\frac{2K_6}{3}\right) J_{n_8}\left(\frac{K_8}{2}\right) J_{n_9}(-2K_9) J_{n_{10}}\left(\frac{2K_{10}}{3}\right) J_{n_{11}}\left(-\frac{2K_{11}}{3}\right) J_{n_{12}}\left(\frac{2K_{12}}{5}\right);$$

Here $J_n(x)$ is the Bessel function.

$\alpha_0(\mathbf{p}-\mathbf{q},\mathbf{p}) = (K_0 + K_7 \cos 2\gamma);$

$\alpha_1 = n_1'+2n_2'+3n_3'+\frac{1}{2}n_4'+n_5'+\frac{3}{2}n_6'+2n_8'-\frac{1}{2}n_9'+\frac{3}{2}n_{10}'-\frac{3}{2}n_{11}'+\frac{5}{2}n_{12}';$

$\alpha_2 = \left(n_1'+2n_2'+3n_3'+\frac{1}{2}n_4'+n_5'+\frac{3}{2}n_6'+2n_8'-\frac{1}{2}n_9'+\frac{3}{2}n_{10}'-\frac{3}{2}n_{11}'+\frac{5}{2}n_{12}'\right) -$

$-\left(n_1+2n_2+3n_3+\frac{1}{2}n_4+n_5+\frac{3}{2}n_6+2n_8-\frac{1}{2}n_9+\frac{3}{2}n_{10}-\frac{3}{2}n_{11}+\frac{5}{2}n_{12}\right);$

$\alpha_3 = n_4'+2n_5'+3n_6'+2n_7'+2n_8'+n_9'+n_{10}'+n_{11}'+n_{12}'-$

$-(n_4+2n_5+3n_6+2n_7+2n_8+n_9+n_{10}+n_{11}+n_{12});$

We apply well-known relation $\exp(ix\sin(y)) = \sum_{n=-\infty}^{\infty} J_n(x)\exp(i n y)$ here and performed a change of variable $t' \to t'+t$. Further we follow a method developed in [15 - 20]. Let us divide our distribution function into high-frequency and low-frequency parts $\tilde{\varphi}(\mathbf{p},t)$ and $\bar{\varphi}(\mathbf{p})$. In first approximation in right-hand parts of equations (8) and (9) we replace full distribution function $\varphi(\mathbf{p},t)$ by $\bar{\varphi}(\mathbf{p})$. So each of equations (8) and

(9) we can change by pair of equations of high-frequency and low-frequency distribution functions. In right part of equation for high-frequency distribution function the terms for which $\alpha_2 \neq 0$ should be received (to right-hand part of equation for low-frequency distribution function the terms corresponded to $\alpha_2 = 0$ should be include, respectively). After integration by $t'$ and taking the real part of result we obtain next equations.

For the scattering on acoustic phonons we obtain

$$\frac{\partial \tilde{\varphi}_{\mathbf{p}}^{ac}(t)}{\partial t} = \sum_{\mathbf{q}} \pi \xi_{a\kappa} \sum_{\substack{n'...n_{12}',\, n_1...n_{12} \\ \alpha_2 \neq 0}} [\{(S_{nn'}(\mathbf{p}-\mathbf{q},\mathbf{p})\cos(\alpha_2 t + \alpha_3 \gamma)\delta(\alpha_0(\mathbf{p}-\mathbf{q},\mathbf{p})+\alpha_1)(\overline{\varphi}_{\mathbf{p}-\mathbf{q}} - \overline{\varphi}_{\mathbf{p}})\} -$$

$$-\{\mathbf{p} \to \mathbf{p}+\mathbf{q}\}] \tag{10}$$

$$0 = \sum_{\mathbf{q}} \pi \xi_{a\kappa} \sum_{\substack{n'...n_{12}',\, n_1...n_{12} \\ \alpha_2 = 0}} [\{(S_{nn'}(\mathbf{p}-\mathbf{q},\mathbf{p})\cos(\alpha_3 \gamma)\delta(\alpha_0(\mathbf{p}-\mathbf{q},\mathbf{p})+\alpha_1)(\overline{\varphi}_{\mathbf{p}-\mathbf{q}} - \overline{\varphi}_{\mathbf{p}})\} -$$

$$-\{\mathbf{p} \to \mathbf{p}+\mathbf{q}\}]. \tag{11}$$

For the scattering on optical phonons we obtain

$$\frac{\partial \tilde{\varphi}_{\mathbf{p}}^{opt}(t)}{\partial t} = \sum_{\mathbf{q}} \pi \xi_{onm} \sum_{\substack{n'...n_{12}',\, n_1...n_{12} \\ \alpha_2 \neq 0}} [\{(S_{nn'}(\mathbf{p}-\mathbf{q},\mathbf{p})\cos(\alpha_2 t + \alpha_3 \gamma)\delta(\alpha_0(\mathbf{p}-\mathbf{q},\mathbf{p})+\alpha_1 + \omega_{\mathbf{q}}')\overline{\varphi}_{\mathbf{p}-\mathbf{q}} -$$

$$- S_{nn'}(\mathbf{p}-\mathbf{q},\mathbf{p})\cos(\alpha_2 t + \alpha_3 \gamma)\delta(\alpha_0(\mathbf{p}-\mathbf{q},\mathbf{p})+\alpha_1 - \omega_{\mathbf{q}}')\overline{\varphi}_{\mathbf{p}})\} - \{\mathbf{p} \to \mathbf{p}+\mathbf{q}\}]; \tag{12}$$

$$0 = \sum_{\mathbf{q}} \pi \xi_{onm} \sum_{\substack{n'...n_{12}',\, n_1...n_{12} \\ \alpha_2 \neq 0}} [\{(S_{nn'}(\mathbf{p}-\mathbf{q},\mathbf{p})\cos(\alpha_3 \gamma)\delta(\alpha_0(\mathbf{p}-\mathbf{q},\mathbf{p})+\alpha_1 + \omega_{\mathbf{q}}')\overline{\varphi}_{\mathbf{p}-\mathbf{q}} -$$

$$- S_{nn'}(\mathbf{p}-\mathbf{q},\mathbf{p})\cos(\alpha_3 \gamma)\delta(\alpha_0(\mathbf{p}-\mathbf{q},\mathbf{p})+\alpha_1 - \omega_{\mathbf{q}}')\overline{\varphi}_{\mathbf{p}})\} - \{\mathbf{p} \to \mathbf{p}+\mathbf{q}\}]; \tag{13}$$

Here we use the well-known relationship $\int_{-\infty}^{0} \exp(i\beta x)dx = \pi\delta(\beta)$.

Further we will consider the limit of weak fields, exactly we suppose that $F_1, F_2 \ll 1$. As it follows from immediate calculations – it is enough to leave the terms with order not more than two by each of quantities $F_1$ and $F_2$ in Taylor expansion of $S_{nn'}$ for construction the first non-vanishing approximation. It is known that the expansion of Bessel function $J_n(x)$ begins from the terms with order $|x|^{|n|}$ so we can limit the values of $n_i$ and $n_i'$ from set $-2,-1,0,1,2$. Thus, constructing of the approximation expressions for right-hand parts of kinetic equations is reduced to selection of same combination of $n_i$ and $n_i'$, which provide the order of the expression not more the second by each of field. For the cases of $b=1$, $b=2$, $b=1/2$ the quantities $S_{nn'}$ contain production of 22 Bessel function, for another value of $b$ this production consists of 24 terms so immediate expansion of that expressions is sufficiently long. To accelerate the calculations Bessel functions are replaced by polynomials contained first three terms of Taylor expansion of this functions. If the order of argument $x$ of Bessel function $J_{n_i}(x)$ by field strengths is equal to 3 or more, the term only with $n_i=0$ is left in expression. If the order of argument is equal to 2, the terms with $n_i=-1,0,1$ are left, and if the order of argument is equal to 1 the terms with $n_i=-2,-1,0,1,2$ are left, respectively. So we receive (in case of $b=1/2$) set of 22 lists each of which contains from one to five Bessel function. To consider all possible cases one should multiply all these lists. The result of the same multiplication is a list contained products of 22 Bessel function which indexes take values from $[-2,2]$ with respect to limitations by argument of Bessel function. But the construction of the same list takes a long time and only little part of it enter in the first non-vanishing approximation. That is why the selection of terms of list is provided by next mean. At the beginning first list of Bessel functions multiplies with the second list and in result list elements with the order not more second by each of fields are left only. Satisfied by our conditions products and corresponded them sets of $n_i$ are entered in two-dimensional lists. On the following step new list of Bessel function is multiplied with received list. Consequently, we receive the list consisted of lists of expanded by field strength expressions $S_{nn'}$ and sets of $n_i$, $n_i'$, corresponded them. Then terms with sets of $n_i$, $n_i'$ correspond to $\alpha_2 \neq 0$ are attached to equation for high-frequency function and other terms are attached to low-frequency equation.

Further we note that quantities $\alpha_0(\mathbf{p}-\mathbf{q},\mathbf{p}) = \varepsilon'(\mathbf{p}-\mathbf{q}) - \varepsilon'(\mathbf{p})$ with accuracy by the second-order terms of $F_1$ and $F_2$, where $\varepsilon'(\mathbf{p}) = \varepsilon(\mathbf{p})/(\hbar\omega_1)$. Speaking strictly, low-frequency distribution functions one should find from functional equations (11) and (13) but with weak field limit we can suppose that low-frequency distribution function is close to equilibrium Boltzmann distribution function $\overline{\varphi}(\mathbf{p}) = \beta\exp(-\varepsilon(\mathbf{p})/(kT))$, where $\beta = \dfrac{16\hbar^2 C_1 \sqrt{\pi}\chi\exp(A/(kT)-\chi^2)}{P^2 L^2(1-\mathrm{erf}(\chi))}$ - is the normalization constant ($\chi = \sqrt{\dfrac{B_1}{4C_1}\dfrac{\hbar\omega_1}{kT}}$, $\mathrm{erf}(x)$ - is error function, $L^2$ - is the square of linear dimension of a basic area of crystal).

By solving the equations (10), (12) we receive high-frequency distribution function for the cases of scattering on acoustic and optical phonons, respectively.

$$\tilde{\varphi}_\mathbf{p}^{ac}(t) = \sum_\mathbf{q} \pi\xi_{ac} \sum_{\substack{n'...n_{12}',\, n_1...n_{12} \\ \alpha_2 \neq 0}} \left( S_{nn'}^1(\mathbf{p}-\mathbf{q},\mathbf{p})\frac{\sin(\alpha_2 t + \alpha_3\gamma)}{\alpha_2}\delta(\varepsilon'(\mathbf{p}-\mathbf{q})-\varepsilon'(\mathbf{p})+\alpha_1)(\overline{\varphi}_{\mathbf{p}-\mathbf{q}}-\overline{\varphi}_\mathbf{p}) - \right.$$

$$\left. - S_{nn'}^1(\mathbf{p},\mathbf{p}+\mathbf{q})\frac{\sin(\alpha_2 t + \alpha_3\gamma)}{\alpha_2}\delta(\varepsilon'(\mathbf{p})-\varepsilon'(\mathbf{p}+\mathbf{q})+\alpha_1)(\overline{\varphi}_\mathbf{p}-\overline{\varphi}_{\mathbf{p}+\mathbf{q}}) \right); \quad (14)$$

$$\tilde{\varphi}_\mathbf{p}^{opt}(t) = \sum_\mathbf{q} \pi\xi_{opt} \sum_{\substack{n'...n_{12}',\, n_1...n_{12} \\ \alpha_2 \neq 0}} \left( S_{nn'}^1(\mathbf{p}-\mathbf{q},\mathbf{p})\frac{\sin(\alpha_2 t + \alpha_3\gamma)}{\alpha_2}\delta(\varepsilon'(\mathbf{p}-\mathbf{q})-\varepsilon'(\mathbf{p})+\alpha_1+\omega_\mathbf{q}')\overline{\varphi}_{\mathbf{p}-\mathbf{q}} - \right.$$

$$- S_{nn'}^1(\mathbf{p}-\mathbf{q},\mathbf{p})\frac{\sin(\alpha_2 t + \alpha_3\gamma)}{\alpha_2}\delta(\varepsilon'(\mathbf{p}-\mathbf{q})-\varepsilon'(\mathbf{p})+\alpha_1-\omega_\mathbf{q}')\overline{\varphi}_\mathbf{p} -$$

$$- \left( S_{nn'}^1(\mathbf{p},\mathbf{p}+\mathbf{q})\frac{\sin(\alpha_2 t + \alpha_3\gamma)}{\alpha_2}\delta(\varepsilon'(\mathbf{p})-\varepsilon'(\mathbf{p}+\mathbf{q})+\alpha_1+\omega_\mathbf{q}')\overline{\varphi}_\mathbf{p} - \right.$$

$$\left.\left. - S_{nn'}^1(\mathbf{p},\mathbf{p}+\mathbf{q})\frac{\sin(\alpha_2 t + \alpha_3\gamma)}{\alpha_2}\delta(\varepsilon'(\mathbf{p})-\varepsilon'(\mathbf{p}+\mathbf{q})+\alpha_1-\omega_\mathbf{q}')\overline{\varphi}_{\mathbf{p}+q} \right)\right); \quad (15)$$

Here $S_{nn'}^1$ are the expanded by fields expressions $S_{nn'}$. Further let us find direct component of current density

$$j_x = en\left\langle \sum_\mathbf{p} \mathrm{v}_x\left(\mathbf{p}+\frac{e}{c}A(t)\right)\tilde{\varphi}(\mathbf{p},t)\right\rangle_t, \quad (16)$$

where angle brackets mean time averaging, $v_x(\mathbf{p}) = \dfrac{\partial \varepsilon(\mathbf{p})}{\partial p_x}$, $n$ - is surface concentration of charge carriers. After averaging we receive next expressions of current density.

$$j_x^{ac} = \pi \xi^{ac} F_1 F_2^2 \sin 2\gamma \sum_{\omega_0'} \left( \sum_{\mathbf{p}} \sum_{\mathbf{q}} G_{\omega_0'}^-(\mathbf{p},\mathbf{q}) \delta(\varepsilon'(\mathbf{p}-\mathbf{q}) - \varepsilon'(\mathbf{p}) + \omega_0')(\overline{\varphi}(\mathbf{p}-\mathbf{q}) - \overline{\varphi}(\mathbf{p})) - \right.$$
$$\left. - \sum_{\mathbf{p}} \sum_{\mathbf{q}} G_{\omega_0'}^+(\mathbf{p},\mathbf{q}) \delta(\varepsilon'(\mathbf{p}) - \varepsilon'(\mathbf{p}+\mathbf{q}) + \omega_0')(\overline{\varphi}(\mathbf{p}) - \overline{\varphi}(\mathbf{p}+\mathbf{q})) \right),$$  (17)

$$j_x^{opt} = \pi \xi^{opt} F_1 F_2^2 \sin 2\gamma \cdot$$
$$\cdot \sum_{\omega_0'} \left( \sum_{\mathbf{p}} \sum_{\mathbf{q}} \left( G_{\omega_0'}^-(\mathbf{p},\mathbf{q}) \delta(\varepsilon'(\mathbf{p}-\mathbf{q}) - \varepsilon'(\mathbf{p}) + \omega_q' + \omega_0') \overline{\varphi}(\mathbf{p}-\mathbf{q}) - \right. \right.$$
$$\left. - G_{\omega_0'}^-(\mathbf{p},\mathbf{q}) \delta(\varepsilon'(\mathbf{p}-\mathbf{q}) - \varepsilon'(\mathbf{p}) - \omega_q' + \omega_0') \overline{\varphi}(\mathbf{p}) \right) -$$  (18)
$$- \sum_{\mathbf{p}} \sum_{\mathbf{q}} \left( G_{\omega_0'}^+(\mathbf{p},\mathbf{q}) \delta(\varepsilon'(\mathbf{p}) - \varepsilon'(\mathbf{p}+\mathbf{q}) + \omega_q' + \omega_0') \overline{\varphi}(\mathbf{p}) - \right.$$
$$\left. \left. - G_{\omega_0'}^+(\mathbf{p},\mathbf{q}) \delta(\varepsilon'(\mathbf{p}) - \varepsilon'(\mathbf{p}+\mathbf{q}) - \omega_q' + \omega_0') \overline{\varphi}(\mathbf{p}+\mathbf{q}) \right) \right);$$

where $\omega_0'$ takes value $-1, -1/2, 0, 1/2, 1$. The expressions $G_{\omega_0'}^-$ and $G_{\omega_0'}^+$ was derived from $S_{nn'}^1(\mathbf{p}-\mathbf{q},\mathbf{p})$ and $S_{nn'}^1(\mathbf{p},\mathbf{p}+\mathbf{q})$, respectively. Let us transform (17) and (18). Firstly, it is clear that term with $\omega_0' = 0$ in (17) don't take part into direct current component. This term corresponds to truly elastic scattering, and from solving the problem with quantum kinetic equation it is immediately followed that this type of scattering doesn't lead to direct current appearance. Further with respect a parity of delta-function we receive from (17), (18):

$$j_x^{ac} = \pi \xi^{ac} F_1 F_2^2 \sin 2\gamma \sum_{\omega_0'} \sum_{\mathbf{p}} \sum_{\mathbf{q}} G_{\omega_0'}(\mathbf{p},\mathbf{q})(\overline{\varphi}(\mathbf{p}-\mathbf{q}) - \overline{\varphi}(\mathbf{p})) \delta(\varepsilon'(\mathbf{p}) - \varepsilon'(\mathbf{p}+\mathbf{q}) + \omega_0'),$$  (19)

where $\omega_0'$ takes values $1/2$ and $1$;

$$j_x^{opt} = \pi \xi^{opt} F_1 F_2^2 \sin 2\gamma \left( \sum_{\mathbf{p}} \sum_{\mathbf{q}} G_0(\mathbf{p},\mathbf{q}) \overline{\varphi}(\mathbf{p}-\mathbf{q}) \delta(\varepsilon'(\mathbf{p}-\mathbf{q}) - \varepsilon(\mathbf{p}) + \omega_q) + \right.$$
$$\left. + \sum_{\omega_0'} \sum_{\mathbf{p}} \sum_{\mathbf{q}} \left( G_{\omega_0'}(\mathbf{p},\mathbf{q}) \overline{\varphi}(\mathbf{p}-\mathbf{q}) \delta(\varepsilon'(\mathbf{p}-\mathbf{q}) - \varepsilon'(\mathbf{p}) + \omega_q' + \omega_0') - G_{\omega_0'}(\mathbf{p},\mathbf{q}) \overline{\varphi}(\mathbf{p}) \delta(\varepsilon'(\mathbf{p}) - \varepsilon'(\mathbf{p}+\mathbf{q}) - \omega_q' + \omega_0') \right) \right),$$
(20)

where $\omega_0'$ takes values $1/2$ and $1$ so.

Here $G_{\omega_0'}(\mathbf{p},\mathbf{q}) = \left( G_{\omega_0'}^-(\mathbf{p},\mathbf{q}) - G_{\omega_0'}^+(\mathbf{p}-\mathbf{q},\mathbf{q}) \right) - \left( G_{-\omega_0'}^-(\mathbf{p}-\mathbf{q},-\mathbf{q}) - G_{-\omega_0'}^+(\mathbf{p},-\mathbf{q}) \right)$.

Quantities $G_{\omega_0'}$ are presented in Appendix.

Expressions (19), (20) are studied numerically. In an ordinary way we change the summation by momentum of electron and phonon to integration by these variables. To remove one integral with delta function we equal argument of delta function to zero and solve numerically an algebraic equation. The remained three integral are taken immediately. Numerical analysis was performed for the case of bilayer graphene ($P = \Delta/v_f$, $\Delta \approx 0{,}2t_\perp$, $D_a = 18 eV$, $D_0 = 1{,}4 \cdot 10^9\, eV/cm$, $T = 50 K$, $\hbar\omega_0 = 0{,}16 eV$, $n = 10^{10}\, cm^{-2}$ [12]). We receive that the direct current component which appear in the case of scattering by acoustic phonons is much less than in the case of scattering by optical phonons ($j^{ac}/j^{opt} \approx 10^{-4}$). The absolute value of current density in the case of electron scattering by optical phonons is $j^{opt} \approx 10^{-2}\, A/cm$, that is supposed accessible [5] at considered temperatures. So the general condition of direct current appearance in the case when on the surface of the sample of material with non-additive energy spectrum two electromagnetic waves with mutually perpendicular planes of polarization are incident is the sufficient inelasticity of scattering. Field dependence of current density in the case of analysis on the base of quantum kinetic equation agrees with the consideration used a Boltzmann kinetic equation with collision term in approximation of constant relaxation time.

The work was done within the Program "Advancement of the higher school science potential". The work was supported by grant of RFBR №10-02-97001-r_povolzhie_a.

# Appendix

The coefficients $K_i$ (7):

$$K_0 = B_1\left(\left((p_x - q_x)^2 + (p_y - q_y)^2\right) - (p_x^2 + p_y^2)\right) + C_1 B_1\left(\left((p_x - q_x)^4 + (p_y - q_y)^4\right) - (p_x^4 + p_y^4)\right) +$$
$$+ C_1 B_1 F_1^2\left(3(p_x - q_x)^2 - 3p_x^2 + (p_y - q_y)^2 - p_y^2\right) + C_1 B_1 F_2^2\left((p_x - q_x)^2 - p_x^2 + 3(p_y - q_y)^2 - 3p_y^2\right);$$

$$K_1 = F_1\left(2B_1 q_x + C_1 B_1\left(3F_1^2 q_x + 2F_2^2 q_x + \right.\right.$$
$$+ 12 p_x^2 q_x + 4 q_x p_y^2 + 4 q_x^3 - 8 p_y q_y q_x + 4 q_x q_y^2 - 24 p_x q_x^2 + 16 p_y q_y p_x - 8 p_x q_y^2\bigg)\bigg);$$

$$K_2 = C_1 B_1 F_1^2\left(3 q_x^2 - 6 p_x q_x - 2 p_y q_y + q_y^2\right);$$

$$K_3 = C_1 B_1 F_1^3 q_x;$$

$$K_4 = F_2\left(2B_1 q_y + C_1 B_1\left(8 p_x q_x p_y - 8 p_x q_x q_y + 4 p_x^2 q_y + 12 p_y^2 q_y - 4 p_y q_x^2 - \right.\right.$$
$$- 12 p_y q_y^2 + 2 F_1^2 q_y + 3 F_2^2 q_y + 4 q_x^2 q_y + 4 q_y^3\bigg)\bigg);$$

$$K_5 = C_1 B_1 F_2^2\left(q_x^2 + 3 q_y^2 - 2 p_x q_x - 3 q_y p_y\right);$$

$$K_6 = C_1 B_1 F_2^3 q_y;$$

$$K_7 = K_8 = C_1 B_1 F_1 F_2^2 q_x;$$

$$K_9 = K_{10} = -4 C_1 B_1 F_1 F_2\left(p_y q_x + p_x q_y - q_x q_y\right);$$

$$K_{11} = K_{12} = C_1 B_1 F_1^2 F_2 q_y;$$

The coefficients $G_i(\mathbf{p}, \mathbf{q})$ (19), (20):

$$G_{1/2}(\mathbf{p}, \mathbf{q}) = B_1^2\left(-16 C_1^2\left(p_y q_x + (p_x - q_x) q_y\right)\left(p_x p_y - (p_x - q_x)(p_y - q_y)\right) +$$
$$+ 2\left(B_1 + 4 B_1 C_1 p_x^2\right)\left(q_y + 2 C_1\left(q_x p_y(2 p_x - q_x) + \left(3 p_y^2 + (p_x - q_x)^2\right) q_y - 3 p_y q_y^2 + q_y^3\right)\right)^2 -$$
$$- 2\left(B_1 + 4 B_1 C_1 (p_x - q_x)^2\right)\left(q_y + 2 C_1\left(-p_y q_x^2 + (2 p_x q_x + 3 p_y q_y)(p_y - q_y) + q_y(p_x^2 + q_x^2 + q_y^2)\right)\right)^2\right);$$

$$G_1(\mathbf{p}, \mathbf{q}) = -\frac{1}{2} B_1^2 C_1 q_x^2 + 4 B_1^3 C_1 q_x^2 q_y(2 p_y - q_y) +$$
$$+ B_1^2 C_1^2\left(\left(p_x^2 - p_y^2 + 2 q_y^2 - 4 p_y q_y\right) q_x^2 - p_x q_y\left(q_y^2 - 10 p_y q_x + 5 q_x q_y\right)\right) +$$
$$+ 16 B_1^3 C_1^2\left(p_y q_x^3(2 p_x - q_x)(p_y - 2 q_y) + 2 p_y q_x^2 q_y(p_x^2 + 2 p_y^2) + 2 p_x^2 q_x q_y^2(p_x - 2 q_x) - q_x^4 q_y^2 - \right.$$
$$- 3 q_y^2 q_x^2(2 p_y^2 - p_x q_x) + 2 p_x p_y q_y^3(p_x - q_x) + 4 p_y q_x^2 q_y^3 + q_y^4\left(p_x q_x - p_x^2 - q_x^2\right)\bigg) +$$

$$+16B_1^3C_1^3\big(4p_xp_y^2q_x^3(p_x^2+p_y^2)-2q_x^4p_y^2(p_y^2-2p_xq_x)-p_y^2q_x^4(6p_x^2+q_x^2)+$$
$$+8p_xp_y^3q_x^2q_y(p_x-2q_x)+6p_yq_x^2q_y(p_y^4-4p_x^3q_x)+2p_x^2p_yq_x^2q_y(11q_x^2+5p_x^2)+20p_y^3q_x^2q_y^3+$$
$$+4p_y^2q_xq_y(2p_yq_x^3+5p_x^3q_y)-2p_yq_x^5q_y(5p_x-q_x)+p_x^4q_xq_y^2(4p_x-15q_x)-q_x^6q_y^2-$$
$$-3p_y^2q_x^2q_y^2(14p_x^2+5p_y^2)+2p_xq_x^3q_y^2(11p_x^2+17p_y^2)-4q_x^4q_y^2(4p_x^2+3p_y^2)+$$
$$+2p_xq_xq_y^2(3q_x^4+q_y^4)+4p_x^3p_yq_y^3(p_x-7q_x)-4p_yq_x^3q_y^3(7p_x-2q_x)-$$
$$-2p_x^2q_y^4(p_x^2+q_y^2+9p_y^2-5p_xq_x)+2p_xp_yq_y^3(p_x-q_x)(5q_y^2+6p_y^2)+q_x^2q_y^5(6p_y-q_y)+$$
$$+2p_xp_yq_xq_y^3(22p_xq_x+9p_yq_y)-q_x^2q_y^4(14p_x^2+15p_y^2)+2q_x^3q_y^4(4p_x-q_x)\big);$$

$$G_0(\mathbf{p},\mathbf{q})=8B_1^3C_1\big(q_xq_y(-2p_yq_x-4p_xq_y+3q_xq_y)-4C_1\big(6p_x^3q_xq_y^2-p_x^2q_y(14q_x^2q_y+q_y^3-$$
$$-2p_y(5q_x^2+q_y^2))-q_x^2(-4p_y^3q_y+3q_y^2(q_x^2+q_y^2)+p_y^2(q_x^2+12q_y^2)-2p_y(2q_x^2q_y+5q_y^3))+$$
$$+p_xq_x(11q_x^2q_y^2+5q_y^4+2p_y^2(q_x^2+6q_y^2)-2p_y(6q_x^2q_y+7q_y^3))\big)-4C_1^2\big(8p_x^5q_xq_y^2+$$
$$+p_x^4q_y(26p_yq_x^2-33q_x^2q_y+4p_yq_y^2-2q_y^3)+2p_x^3q_x(2p_y^2(5q_x^2+11q_y^2)-$$
$$-2p_y(18q_x^2q_y+13q_y^3)+9(3q_x^2q_y^2+q_y^4))-q_x^2(-6p_y^5q_y+3q_y^2(q_x^2+q_y^2)^2+p_y^4(2q_x^2+33q_y^2)-$$
$$-4p_y^3(5q_x^2q_y+14q_y^3)-6p_yq_y(q_x^4+4q_x^2q_y^2+3q_y^4)+3p_y^2(q_x^4+12q_x^2q_y^2+15q_y^4))+$$
$$+2p_x^2(p_y^3(28q_x^2q_y+6q_y^3)-q_y^2(22q_x^4+17q_x^2q_y^2+q_y^4)-3p_y^2(5q_x^4+25q_x^2q_y^2+3q_y^4)+$$
$$+p_y(37q_x^4q_y+60q_x^2q_y^3+5q_y^5))+2p_xq_x(2p_y^4(q_x^2+9q_y^2)-2p_y^3(16q_x^2q_y+21q_y^3)+$$
$$+p_y^2(8q_x^4+65q_x^2q_y^2+39q_y^4)-p_y(17q_x^4q_y+46q_x^2q_y^3+17q_y^5)+3(q_x^4q_y^2+4q_x^2q_y^4+q_y^6))\big)\big);$$